\documentclass[preprint,showpacs,preprintnumbers,amsmath,amssymb,prb]{revtex4}

\usepackage{graphicx}
\usepackage{dcolumn}
\usepackage{bm}

\begin{document}
\title
{
{\it In-situ} epitaxial growth of superconducting La-based bilayer cuprate 
thin films
}
\author{I. Tsukada}
\email{ichiro@criepi.denken.or.jp} 
\author{Y. Nagao}
\altaffiliation[Also at ]{Department of Physics, Tokyo Universtiy of Science, 
Kagurazaka, Tokyo 162-8601, JAPAN}

\author{Yoichi Ando}
\altaffiliation[Also at ]{Department of Physics, Tokyo Universtiy of Science, 
Kagurazaka, Tokyo 162-8601, JAPAN}

\affiliation{
Central Research Institute of Electric Power Industry, 
2-11-1 Iwado-kita, Komae, Tokyo 201-8511, JAPAN
}

\date{\today}

\begin{abstract}
We investigate the epitaxial growth of bilayer cuprate 
La${}_2$CaCu${}_2$O${}_{6+\delta}$ using pure ozone as an oxidant, 
and find that even the crystal with parent composition without cation substitution 
can show metallic behavior with the aid of epitaxial strain effect. 
The hole concentration is controlled simply by excess-oxygen doping, 
and the films grown under the optimum conditions exhibit superconductivity 
below 30~K. 
This is the first result on the superconductivity of bilayer 
La${}_2$CaCu${}_2$O${}_{6+\delta}$ induced purely by the excess oxygen. 
\end{abstract}
\pacs{74.62.Bf,74.62.Dh,74.72.Dn,74.78.Bz}

\maketitle


Epitaxial growth technique is quite useful for the study of 
high-$T_c$ superconductors, because one can easily 
modify lattice parameters by choosing an appropriate substrate. 
In particular, La${}_{2-x}$Sr${}_x$CuO${}_4$ (LSCO) thin films can 
have higher $T_c$ than bulk crystals by adding in-plane compressive strain.
\cite{Sato1,Locquet1} 
On the contrary to such a monolayer cuprate as LSCO, few results have been 
reported on the role of epitaxial strain in bilayer cuprates 
YBa${}_2$Cu${}_3$O${}_7$ (YBCO) or 
Bi${}_2$Sr${}_2$CaCu${}_2$O${}_8$ (BSCCO) so far, 
probably because no remarkable improvement of superconducting properties 
has been found in these compounds; 
Cu-O chain layers in YBCO or cleavable Bi-O layers in BSCCO 
may relax rapidly the lattice mismatch from the substrate in the initial 
growth stage. 
However, since bilayer cuprates show in general higher $T_c$ than monolayer 
ones, it would be fascinating if a remarkable $T_c$ enhancement is 
induced by the epitaxial strain. 
La${}_{2-x}$(Sr,Ca)${}_{1+x}$Cu${}_2$O${}_{6+\delta}$ (La2126) is a good 
bilayer compound for the study of epitaxial growth, 
because this compound has a simpler crystal structure than YBCO and BSCCO. 
A few results on thin-film growth of La2126 on SrTiO${}_3$ (100) 
have been reported,
\cite{Nakamura1,Verbist1} 
but unfortunately superconducting films have never been obtained. 
As is widely known, La2126-type structure can be rather easily obtained 
by a conventional solid-state reaction method,
\cite{Torrance1,Tamegai1} 
although post annealing under the high-pressure oxygen of 10$\sim$400~atm 
is indispensable to induce superconductivity. 
\cite{Cava1,Kinoshita1} 
The results of Refs.~3 and 4 imply that their oxidation techniques were 
not efficient enough to oxidize the films as high-pressure oxygen does. 
Thus, one needs to develop a special technique to oxidize samples. 
One of the good candidates for oxidizing gas is pure ozone, 
because its handling techniques have been well established for more than 
a decade,
\cite{Berkley1,Nakayama1}
and have recently been applied to synthesize new superconductors 
such as Ba${}_2$CuO${}_{4-\delta}$ and Sr${}_2$CuO${}_{4-\delta}$.
\cite{Yamamoto1,Karimoto1} 
It has been reported that chemical activity of ozone is higher than 
that of molecular oxygen at GPa pressure,
\cite{Suzuki1}
although the life time of ozone at high temperatures is quite short 
({\it e.g.}, $\sim$10${}^0$ s at 200${}^{\circ}$C, 
and $\sim$10${}^{-3}$ s at 500${}^{\circ}$C).
\cite{Horvath1} 
In this paper we report a successful growth of superconducting La2126 
thin films using pure ozone. 
In order to see the sole effect of oxidation by ozone, 
we concentrate on the parent composition, La${}_2$CaCu${}_2$O${}_6$ (LCCO), 
in this report. 
A more-detailed account of the results for Sr and Ca substituted compounds 
will be described elsewhere.

All the films were prepared by pulsed laser deposition (PLD). 
A non-superconducting polycrystalline target with the nominal composition 
of La:Ca:Cu = 2:1:2 was prepared by a conventional solid-state 
reaction method. 
The target was ablated by KrF excimer laser ($\lambda$ = 248~nm) 
under pure ozone. 
It should be noted that we can access higher gas-pressure regime 
in a PLD chamber than in a conventional molecular-beam epitaxy (MBE) chamber. 
The ozone pressure for the present experiment was set to 15~mPa, 
which is higher by one order of magnitude than the typical pressure 
in MBE experiments. 
To control the oxygen contents, we change the temperature at which 
ozone is turned off ($T_{O_3}$) during the cooling-down process. 
The substrate temperature during the growth is kept at 800${}^{\circ}$C 
for all the films shown in this paper. 
The resistivity and Hall coefficient measurements are done for samples 
mechanically patterned in a six-terminal shape in order not to add 
any heat-treatment process that might change the oxygen contents of the sample.

\begin{figure}
\includegraphics[width=100mm]{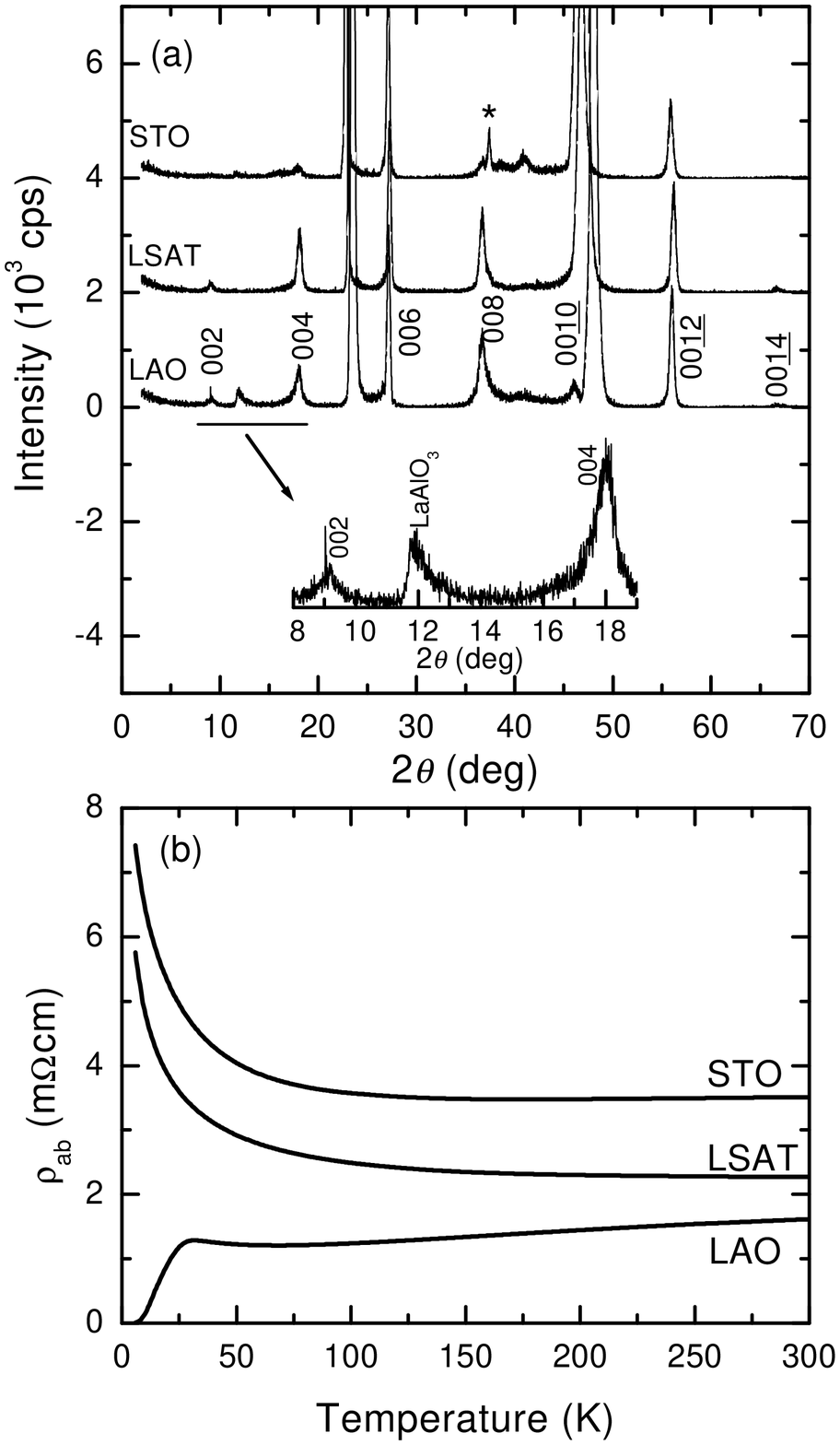}
\caption{
(a) X-ray diffractions of La${}_2$CaCu${}_2$O${}_{6+\delta}$ thin films 
grown on LaAlO${}_3$ (100), LSAT (100) and SrTiO${}_3$ (100); 
the diffraction peak from (La,Ca)${}_2$CuO${}_4$ impurity phase is indicated 
by $*$. 
Expanded diffraction pattern between 
8${}^{\circ}$$<$2${\theta}$$<$18${}^{\circ}$ is also shown for the film 
grown on LaAlO${}_3$. 
(b) Temperature dependences of the resistivity of the corresponding films.
}
\label{Fig.1}
\end{figure}

First, we tried to determine what substrate is suitable for LCCO thin films. 
Figure~\ref{Fig.1}(a) shows the x-ray (Cu$K_{\alpha}$) diffraction of LCCO 
films grown on LaAlO${}_3$ (100),
\cite{LAO} 
(LaAlO${}_3$)${}_{0.3}$(SrAl${}_{0.5}$Ta${}_{0.5}$O${}_3$)${}_{0.7}$ (100), 
and SrTiO${}_3$ (100). 
(We will call them LAO, LSAT, and STO, respectively.) 
These films are fully oxidized with $T_{O_3}$ = 65${}^{\circ}$C. 
The films grown on LAO and LSAT show highly $c$-axis oriented structure 
of La2126 phase. 
Both films are free from an impurity phase like (La,Ca)${}_2$CuO${}_4$; 
the 002 peak of (La,Ca)${}_2$CuO${}_4$ are expected to show up at 
13.5${}^{\circ}$$<$2${\theta}<$14${}^{\circ}$,
\cite{Sato1} 
which is completely absent in the films grown on LSAT and LAO. 
In contrast, the film grown on STO shows weaker diffraction intensities 
and includes (La,Ca)${}_2$CuO${}_4$ as an impurity phase (indicated by $*$), 
which indicates that the lattice misfit is too large in this case, and 
the thermodynamically stable (La,Ca)${}_2$CuO${}_4$ may show up.
Thus, the STO substrate, which was used in the previous experiments,
\cite{Nakamura1,Verbist1} 
is considered to be inappropriate for the growth of LCCO thin films. 
The $c$-axis length of the film on LAO is estimated to be $\approx$ 19.65{\AA}, 
which is longer than that of the film on LSAT (19.62{\AA}). 
These results can be explained as the result of lattice mismatch: 
the misfit of LCCO is positive to LAO ($\approx$ 1.20{\%}) and 
negative to LSAT ($\approx$ -0.89{\%}). 
In this situation, it is expected that the LCCO film feels 
a compressive strain on LAO, while it feels a tensile one on LSAT; 
consequently, the film on LAO has longer $c$-axis length than that on LSAT. 
Our results are consistent with this picture, and therefore, we conclude 
that the epitaxial starin effect works in LCCO films on LAO and LSAT. 
It should be noted that the $c$-axis lengths of both films are larger 
than that reported for polycrystalline sample La${}_2$CaCu${}_2$O${}_{6.037}$, 
19.5169{\AA}, that were prepared under 1~atm oxygen.
\cite{Fuertes1} 
This deviation is probably due to the difference 
in the amount of excess oxygen, 
and suggests that the ozone annealing is really effective in giving 
large amount of excess oxygen.

The substrate dependence appears also in the temperature dependence of 
the in-plane resistivity $\rho_{ab}$(T) as shown in Fig.~\ref{Fig.1}(b). 
The film on STO shows semiconducting behavior as is easily expected 
from its poor crystallization suggested from the x-ray diffraction. 
What is interesting is that the films on LSAT and LAO exhibit contrasting 
behavior; the former shows semiconducting behavior down to 2.5~K, 
while the latter keeps metallic behavior and exhibits superconductivity 
below $\approx$ 30~K. 
The room-temperature resistivity of the superconducting sample 
(2~m${\Omega}$cm) is lower than that reported for single crystals.
\cite{Okuya1} 
However, we could not reduce the resistivity below 1~m$\Omega$cm as other 
typical superconducting cuprates, 
which might be due to oxygen inhomobeneity in the sample. 
The observed substrate dependence is probably due to the directions 
of strain as was observed in LSCO thin films;
\cite{Sato1,Locquet1} 
the compressive strain is helpful in enhancing metallic behavior 
while the tensile strain gives the opposite effect. 
However, it is not so simple to understand from a microscopic vewpoint 
that similar epitaxial-strain effects are observed both 
in monolayer LSCO and in bilayer LCCO: 
In LSCO, the characteristic tilting of the CuO${}_6$ octahedra, 
which is believed to dominate $T_c$, can be controlled by the epitaxial strain. 
On th other hand, such a characteristic structure is absent in LCCO, 
and we need further studies on structural analysis.

\begin{figure}
\includegraphics[width=100mm]{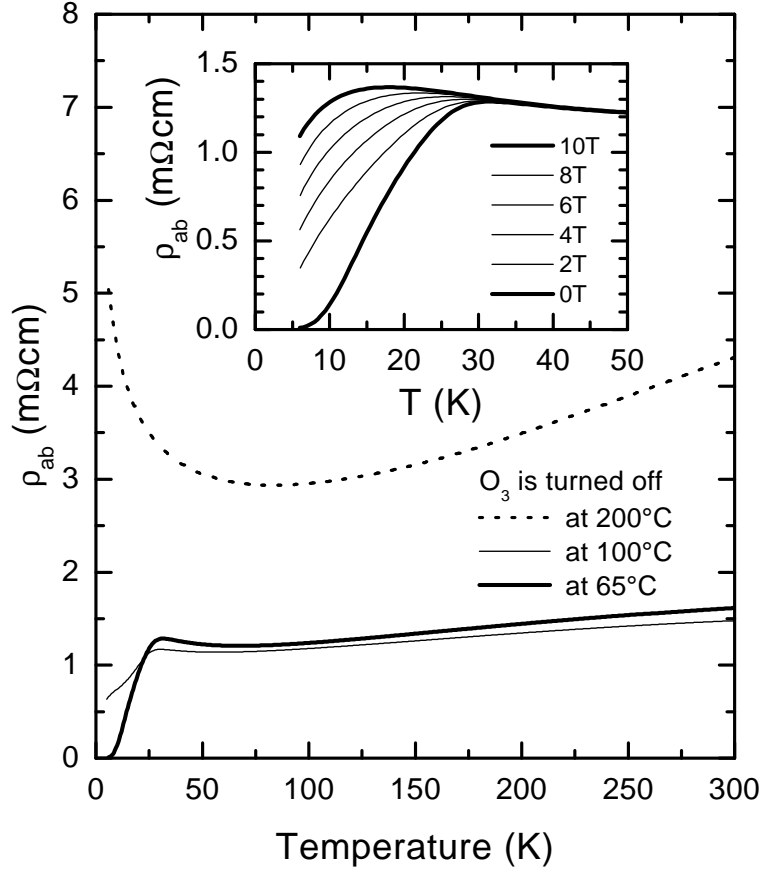}
\caption{
Temperature dependences of the resistvitiy of La${}_2$CaCu${}_2$O${}_{6+\delta}$ 
thin films prepared at different $T_{\rm O_3}$. 
The film prepared with $T_{\rm O_3}$ = 200${}^{\circ}$C shows an insulating 
bahabior, while those prepared with $T_{\rm O_3}$ = 100${}^{\circ}$C and 
65${}^{\circ}$C show superconductivity below $\approx$30~K.
}
\label{Fig.2}
\end{figure}

According to the above results, we chose the LAO substrate for further study, 
in which we tried to control the amount of extra oxygen. 
Figure~\ref{Fig.2} shows the $T_{\rm O_3}$ dependence of $\rho_{ab}(T)$. 
If we turn off ozone at $T_{\rm O_3}$ = 200${}^{\circ}$C, 
which is typical for the growth of LSCO films with ozone as an oxidant, 
superconductivity does not appear. 
$\rho_{ab}(T)$ exhibits metallic behavior down to $\approx$~80~K, but 
resistivity upturn appears in the lower-temperature region. 
This temperature dependence is similar to that reported for 
La${}_{2-x}$Ca${}_{1+x}$Cu${}_2$O${}_{6+\delta}$ ceramic samples 
annealed under 1$\sim$2~atm oxygen,
\cite{Kinoshita1,Kinoshita2} 
but is more metallic than that of the as-grown bulk crystals of 
La${}_{1.90}$Ca${}_{1.10}$Cu${}_2$O${}_6$.
\cite{Okuya1} 
By decreasing $T_{\rm O_3}$ to 100${}^{\circ}$C, 
$\rho_{ab}(T)$ decreases significantly and superconductivity shows up; 
the room-temperature resistivity decreases roughly by three times, 
and the resistivity drop appears around 30~K. 
However, this film does not show zero resistivity above 2.5~K, which suggests 
that the superconducting regions are segmented. 
Only after reducing $T_{\rm O_3}$ to 65${}^{\circ}$C, 
we could obtain zero resistivity at $\approx$ 5~K. 
We emphasize that this is the first result to observe zero resistivity in LCCO. 
Although Fuertes {\it et al.}
\cite{Fuertes1} 
have reported a slight decrease of AC magnetic 
susceptibility below 45~K for ceramic LCCO, they observed neither metallic 
nor superconducting behavior in resistivity measurements, 
and bulk superconductivity was not confirmed. 
In order to confirm the presence of bulk superconductivity, 
we have applied magnetic field along the $c$ axis 
(inset of Fig.~\ref{Fig.2}), 
and have observed a typical $T_c$ suppression by the magnetic field. 
Superconductivity remains even at $H$ = 10~T, 
which may reject the possibility of filamentary superconductivity. 
It should be noted that the difference between the films prepared at 
$T_{\rm O_3}$ = 65${}^{\circ}$C and 100${}^{\circ}$C is found only 
in zero-resistivity temperature, 
while the magnitude of the normal-state resistivity and the $T_c$ onset are 
almost unchanged. 
Therefore, decreasing $T_{\rm O_3}$ from 100${}^{\circ}$C 
to 65${}^{\circ}$C does not seem to change the carrier density drastically. 
One possibility is that by lowering $T_{\rm O_3}$ the distribution of 
extra oxygen becomes more homogeneous, 
but this should be confirmed experimentally.

\begin{figure}
\includegraphics[width=100mm]{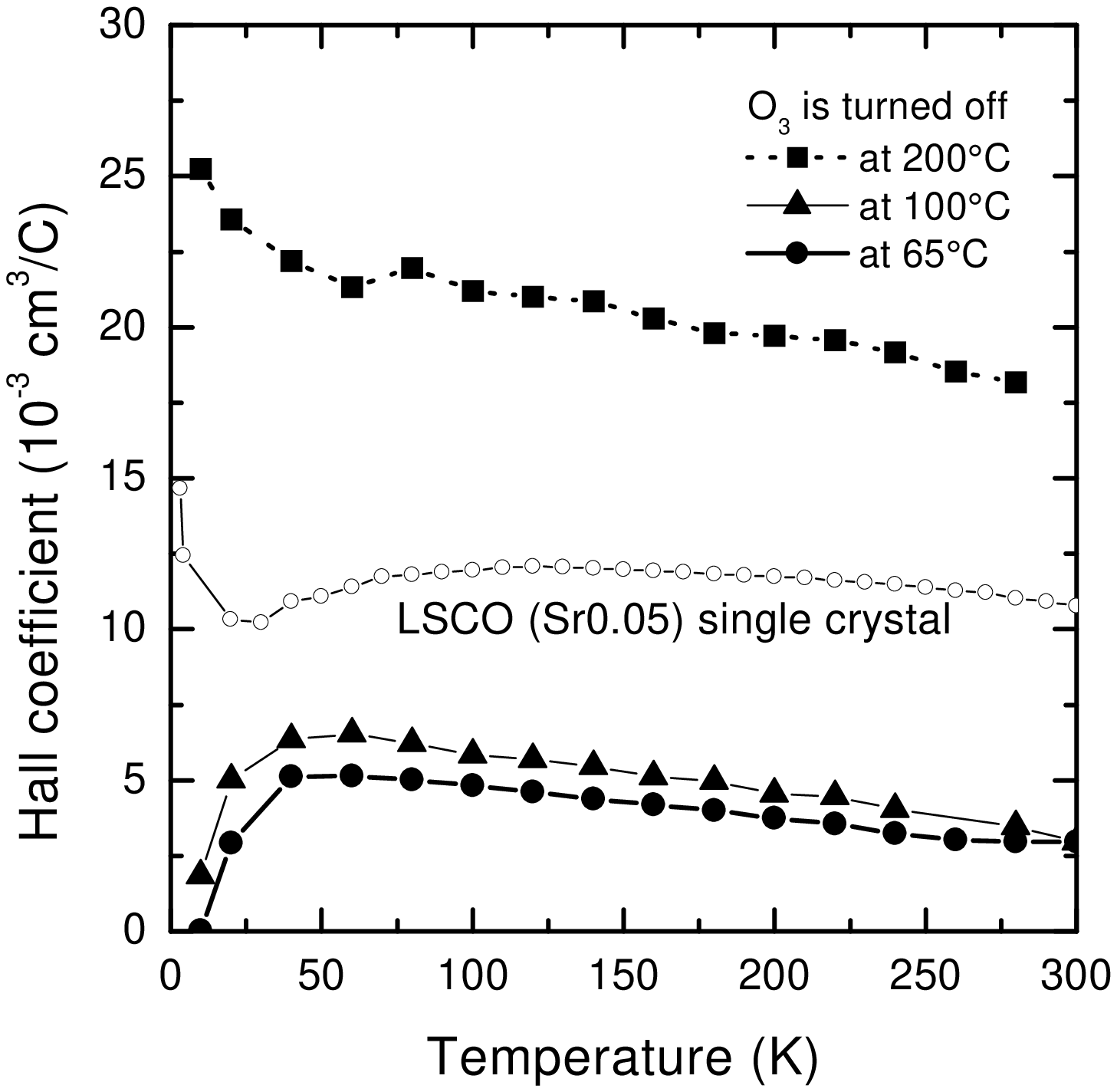}
\caption{
Temperature dependences of the Hall coefficient of 
La${}_2$CaCu${}_2$O${}_{6+\delta}$ films. 
The Hall coefficient of the insulating film prepared 
at $T_{\rm O_3}$ = 200${}^{\circ}$C exceeds 1$\times$10${}^{-2}$ cm${}^3$/C 
at room temperature, while those of superconducting films are 
$\approx$3$\times$10${}^{-3}$ cm${}^3$/C at room temperature. 
The data of single-crystal La${}_{1.95}$Sr${}_{0.05}$CuO${}_4$ is taken from 
Ref. 18; 
this gives an approximate boundary of the carrier concentration 
between insulating and superconducting phases for LSCO.
}
\label{Fig.3}
\end{figure}

The change of carrier concentration is confirmed more directly by 
a Hall-effect measurement, where we found the results being consistent 
with the behavior of $\rho_{ab}(T)$. 
Figure~\ref{Fig.3} shows the temperature dependence of the Hall coefficient 
$R_H$ of the films shown in Fig.~\ref{Fig.2}. 
The film prepared at $T_{\rm O_3}$ = 200${}^{\circ}$C seems to have 
insufficient amount of holes; 
$R_H$ exceeds 1$\times$10${}^{-2}$~cm${}^3$/$C$. 
This value is typical for LSCO with $x$ = 0.05, which is non-superconducting.
\cite{Ando1} 
Once we decrease $T_{\rm O_3}$ to 100${}^{\circ}$C, 
$R_H$ remarkably decreases to the order of 10${}^{-3}$~cm${}^3$/$C$. 
The films prepared at $T_{\rm O_3}$ = 100${}^{\circ}$C and 65${}^{\circ}$C 
have almost the same $R_H$ 
$\approx$ 3$\times$10${}^{-3}$~cm${}^3$/$C$ at room temperatures, 
and no remarkable difference is found between them down to 20~K. 
This is consistent with the resistivity data. 
The rapid decrease at $T <$ 40~K is due to the superconducting transition.

We can roughly evaluate the strength of oxidation 
by comparing the measured $R_H$ with reported values. 
Nishikawa {\it et al.}
\cite{Nishikawa1} 
reported the Hall coefficient for a polycrystalline 
La${}_2$CaCu${}_2$O${}_6$ sample annealed under 20~atm oxygen. 
They reported $R_H$ $\approx$ 7$\times$10${}^{-3}$~cm${}^3$/$C$ 
at room temperature, which is higher than that of our films 
prepared at $T_{\rm O_3}$ $\leq$ 100${}^{\circ}$C. 
This implies that the ozone annealing down to 100${}^{\circ}$C or lower 
is more effective than the annealing in 20~atm oxygen.

Now let us discuss the role of excess oxygen on hole doping. 
The first report on superconductivity in La2126 gave us an impression 
that the Sr substitution of La was indispensable for superconductivity,
\cite{Cava1} 
and more detailed systematic studies on 
La${}_{2-x}$Ca${}_{1+x}$Cu${}_2$O${}_{6+\delta}$ 
also demonstrated that $T_c$ increased with increasing $x$ when 
annealed under 10 $\sim$ 20~atm oxygen.
\cite{Kinoshita1} 
In the latter case, however, $T_c$ and the magnitude of resistivity 
seem to approach certain saturation values upon increasing 
the annealing pressure to 400~atm, 
which implies that the obvious $x$ dependences are diminished when the 
excess oxygen is fully added. 
Similar results were obtained for single crystals annealed in 400~atm oxygen.
\cite{Okuya1} 
Therefore, we had better conclude that the Ca or Sr concentration is not a good 
measure of hole density, 
but rather a complicated combination of alkaline-earth substitution 
and high-pressure annealing determine $T_c$ and the transport properties 
in La2126 compounds.
\cite{Liu1} 
This consideration was supported by a Hall-effect measurement;
\cite{Nishikawa1} 
it was reported that $R_H$ $\approx$ 7$\times$10${}^{-3}$~cm${}^3$/$C$ 
for pure La${}_2$CaCu${}_2$O${}_6$ was suppressed only to 
$R_H$ $\approx$ 4.5$\times$10${}^{-3}$~cm${}^3$/$C$ for 
La${}_{1.2}$Ca${}_{0.8}$Cu${}_2$O${}_6$. 
This change is far smaller than that expected from the Ca concentration 
by using the simple formula $R_H$ = 1/($ne$), where $n$ and $e$ are 
hole density and electric charge, respectively. 
It is likely that a non-negligible number of holes doped by Ca substitution 
is compensated by a decrease of excess oxygen, and the measured Hall 
coefficient becomes far larger than that expected from the Ca concentration.

In the present case, the hole density is simply dominated by the amount of 
excess oxygen, which will be useful to establish 
a hole-density vs $T_c$ phase diagram for La2126. 
As is suggested by Fig.~\ref{Fig.3}, we can use the Hall coefficient 
as a measure of hole concentration, 
even though the exact amount of excess oxygen remains unknown. 
In that sense, our results are similar to those reported 
for La${}_{1.89}$Ca${}_{1.11}$Cu${}_2$O${}_{6+\delta}$ single crystals,
\cite{Watanabe2} 
where the authors investigated in detail the annealing pressure dependence 
of the resistivity and the Hall coefficient. 
Our advantage over the bulk crystals is the absence of cation substitution 
that might introduce crystallographic disorder. 
If we compare the Hall coefficient at room temperature between 
La${}_{1.89}$Ca${}_{1.11}$Cu${}_2$O${}_{6+\delta}$ single crystals 
\cite{Watanabe2} 
and our films, 
the hole concentration of our films prepared with 
$T_{\rm O_3}$ = 65${}^{\circ}$C is almost the same as that of 
the single crystals annealed at 300~atm oxygen. 
This indicates that the ozone annealing in the vacuum chamber is comparable 
to the high-pressure oxygen of 100~atm order. 
For bulk samples, 400~atm might be the highest oxygen annealing pressure 
by HIP, but this pressure seems to be still insufficient 
to obtain the optimally doped samples. 
On the other hand, the ozone pressure for this study is set to 15~mPa, 
but this is not the upper limit. 
It is technically possible to increase the ozone pressure higher than 15~mPa, 
which seems to be the only way to obtain not only to 
the optimally-doped but also to the over-doped samples of La2126 compounds.

Finally, we breifly mention where the excess oxygen goes. 
It was confirmed for La${}_{1.82}$Ca${}_{1.18}$Cu${}_2$O${}_{6+\delta}$ 
(annealed at 400~atm oxygen) by neutron diffraction
\cite{Kinoshita3} 
that the excess oxygen is located between two neighboring CuO${}_2$ planes. 
Thus, it is natural to expect that the excess oxygen is located 
at this position in our films also. 
However, we cannot believe that all of the excess oxygen occupy this position. 
According to Ref.~23, the $c$-axis length of the sample with excess oxygen is 
longer only by $\approx$~0.005{\AA} than that of the sample 
without excess oxygen. 
In contrast, our LCCO films have far longer $c$-axis length regardless of 
substrates compared to the bulk LCCO samples;
\cite{Fuertes1} 
even the $c$-axis length of the film on LSAT, which feels an in-plane 
tensile strain, is longer by $\approx$~0.11{\AA} than that of the bulk samples. 
Thus, it is difficult to attribute the observed remarkbale elongation 
of the $c$ axis only to the oxygen between the neighboring CuO${}_2$ planes. 
It is rather likely that the excess oxygen goes also into the interstitial 
sites at La${}_2$O${}_2$ blocks like in La${}_2$CuO${}_{4+\delta}$; 
the elongation of the $c$-axis length by 0.04 $\sim$ 0.06~{\AA} was 
reported for La${}_2$CuO${}_{4+\delta}$,
\cite{Jorgensen1} 
which suggests that the remarkbale elongation of the $c$-axis length 
in the present case is due to the same mechanism.

In any case, the ozone-annealed LCCO parent compound provides a unique 
opportunity for the study of bilayer cuprates, where the carrier concentration 
can be widely controlled without cation substitution. 
It was discussed that the cation substitution introduces ``quenched'' disorder 
while the oxygen intercalation gives ``annealed'' disorder,
\cite{Wells1} 
and the latter seems to give much benign influenece on the structure 
of CuO${}_2$ planes. 
Such a situation has never been realized in bilayer YBCO, 
because the change in oxygen contents is always accompanied by the change 
in transport properties of Cu-O chain layer, and thus, 
one cannot purely observe the response of CuO${}_2$ bilayers. 
We can tune the transport properties of the LCCO films in a wide range 
from insulator to superconductor, which provides a more ideal situation 
than the extensively studied YBCO.

In summary, we have succeeded in growing superconducting 
La${}_2$CaCu${}_2$O${}_{6+\delta}$ thin films by choosing an appropriate 
substrate and by performing a strong oxidation using pure ozone. 
As was reported for monolayer cuprate films, the slight compressive strain 
along the in-plane direction is found to be helpful 
to make La${}_2$CaCu${}_2$O${}_{6+\delta}$ metallic. 
The low-temperature ozone annealing is found to be comparable to 
or more effective than high-pressure oxygen annealing. 
Our next step is to extend target compounds to multilayer cuprates where 
the combination of epitaxial strain and the strong oxidation will become 
more important to obtain higher $T_c$.

We thank S. Komiya for fruitful discussions.

\newpage

\begin{references}
\bibitem{Sato1} H. Sato and M. Naito, 
Physica C {\bf 274}, 221 (1997); H. Sato, A. Tsukada, M. Naito, and A. Matsuda, 
Phys. Rev. B {\bf 61}, 12447 (2000). 
\bibitem{Locquet1} J. -P. Locquet, J. Perret, J. Fompeyrine, 
E. M{\"a}chler, J. W. Seo, and G. Van Tendeloo, 
Nature (London) {\bf 394}, 453 (1998). 
\bibitem{Nakamura1} K. Nakamura, H. Nobumasa, K. Shimizu, and T. Kawai, 
Physica C {\bf 221}, 387 (1994). 
\bibitem{Verbist1} K. Verbist, O. Milat, G. Van Tendeloo, F. Arrouy, 
E. J. Williams, C. Rossel, E. M\"achler, and J. -P. Locquet, 
Phys. Rev. B {\bf 56}, 853 (1997). 
\bibitem{Torrance1} J. B. Torrance, Y. Tokura, A. Nazzal, 
and S. S. P. Parkin, 
Phys. Rev. Lett. {\bf 60}, 542 (1988). 
\bibitem{Tamegai1} T. Tamegai and Y. Iye, 
Physica C {\bf 159}, 181 (1989). 
\bibitem{Cava1} R. J. Cava, B. Batlogg, R. B. van Dover, J. J. Krajewski, 
J. V. Waszczak, R. M. Fleming, W. F. Oeck Jr., L. W. Rupp Jr., 
P. Marsh, A. C. W. P. James, and L. F. Schneemeyer, 
Nature (London) {\bf 345}, 602 (1990). 
\bibitem{Kinoshita1} K. Kinoshita, H. Shibata, and T. Yamada, 
Physica C {\bf 171}, 523 (1990); 
K. Kinoshita and T. Yamada, 
Phys. Rev. B {\bf 46}, 9116 (1992). 
\bibitem{Berkley1} D. D. Berkley, B. R. Johnson, N. Anand, K. M. Beachamp, 
L. E. Cpmrpy, A. M. Goldman, J. Maps, K. Mauersberger, M. L. Mecartney, 
J. Morton, M. Tuominen, and Y. -J. Zhang, 
Appl. Phys. Lett. {\bf 53}, 572 (1988). 
\bibitem{Nakayama1} Y. Nakayama, I. Tsukada, and K. Uchinokura, 
J. Appl. Phys. {\bf 70}, 4371 (1991). 
\bibitem{Yamamoto1} H. Yamamoto, M. Naito, and H. Sato, 
Physica C {\bf 38}, 29 (2000). 
\bibitem{Karimoto1} S. Karimoto, H. Yamamoto, T. Greibe, and M. Naito, 
Jpn. J. Appl. Phys. {\bf 40}, L127 (2001). 
\bibitem{Suzuki1} R. O. Suzuki, T. Ogawa, and K. Ono, 
J. Amer. Ceram. Soc. {\bf 82}, 2033 (1998). 
\bibitem{Horvath1} M. Horvath, L. Bilitzky, and J. Huettner, 
{\it Ozone}, Elesevier Science Publishers, Amsterdam, The Netherland, (1985). 
\bibitem{LAO} To say excatly, LaAlO${}_3$ has a rhombohedral symmetry. 
However, we follow the pseudo-cubic notation in this paper 
to make the comparison with other substrates easier. 
\bibitem{Fuertes1} A. Fuertes, X. Obradors, J.M. Navarro, P. Gomez-Romero, 
N. Casa{\~n}-Pastor, F. P\'erez, J. Fontcuberta, C. Miravitlles, 
J. Rodoriguez-Carvajal, and B. Mart{\'i}nez, 
Physica C {\bf 170}, 153 (1990).
\bibitem{Kinoshita2} 
K. Kinoshita, H. Shibata, and T. Yamada, 
Physica C {\bf 176}, 433 (1991). 
\bibitem{Ando1} Y. Ando, Y. Kurita, S. Komiya, S. Ono, and K. Segawa, 
unpublished. 
Note that the number of CuO${}_2$ planes per primitive-cell-thick slab 
is different between LSCO and LCCO. 
If one considers the density of holes per CuO${}_2$ plane, 
$R_H$ of LCCO multiplied by factor $\approx$~1.34 should be compared with 
$R_H$ of LSCO, 
but this modification does not change the present discussion. 
\bibitem{Nishikawa1} T. Nishikawa, S. Shamoto, M. Sera, M. Sato, 
S. Ohsugi, Y. Kitaoka, and K. Asayama, 
Physica C {\bf 209}, 553, (1993). 
\bibitem{Okuya1} M. Okuya, T. Kimura, R. Kobayashi, J. Shimoyama, K. Kitazawa, 
K. Yamafuji, K. Kishio, K. Kinoshita, and T. Yamada, 
J. Superconduct. {\bf 7}, 313 (1994). 
\bibitem{Liu1} H. B. Liu, D. E. Morris, A. P. B. Sinha, and X. X. Tang, 
Physica C {\bf 174}, 28 (1991). 
\bibitem{Watanabe2} T. Watanabe, K. Kinoshita, and A. Matsuda, 
Phys. Rev. B {\bf 47}, 11544 (1993). 
\bibitem{Kinoshita3} K. Kinoshita, F. Izumi, T. Yamada, and H. Asano, 
Phys. Rev. B {\bf 45}, 5558 (1992). 
\bibitem{Jorgensen1} J. D. Jorgensen, B. Dabrowski, S. Pei, D. G. Hinks, 
L. Soderholm, B. Morosin, J. E. Schirber, E. L. Venturini, and D. S. Ginley, 
Phys. Rev. B {\bf 38}, 11337 (1988). 
\bibitem{Wells1} B. O. Wells, Y. S. Lee, M. A. Kastner, R. J. Christianson, 
R. J. Birgeneau, K. Yamada, Y. Endoh, and G. Shirane, 
Science {\bf 277}, 1067 (1997). 
\end{references}
\end{document}